\documentclass[slac_one]{revtex4}
\usepackage{graphicx}
\usepackage{fancyhdr}
\newcommand {\be}{\begin{equation}}
\newcommand {\ee}{\end{equation}}
\newcommand {\ba}{\begin{eqnarray}}
\newcommand {\ea}{\end{eqnarray}}
\begin{document}

\title{Combined  Analysis of Electric Dipole Moments and Lepton Flavor Violating Rare Decays
\footnote{Presented at ICHEP 08 by Y. Farzan as the IUPAP young scientist prize ceremony talk}} 

%

\author{Y. Ayazi and Y. Farzan}
\affiliation{ Institute for research in fundamental sciences
(IPM), P.O. Box 19395-5531, Tehran, Iran}

\begin{abstract}
In the context of general Minimal Supersymmetric Standard Model
(MSSM), new sources for Lepton Flavor Violation (LFV) as well as
CP-violation appear. We show that in the presence of LFV sources,
the electric dipole moment of the electron ($d_e$) can receive new
contributions. In particular, $d_e$ can receive a significant
contribution at one loop level from the phase of the trilinear
$A$-term of the staus, $\phi_{A_\tau}$. We discuss how we can
derive information on $\phi_{A_\tau}$ by combining  the
information on $d_e$ with that on the LFV  decay modes of the
$\tau$ lepton. We then discuss if this approach can be considered
as an alternative to the direct  measurement of $\phi_{A_\tau}$ at
ILC.
\end{abstract}

\maketitle

\thispagestyle{fancy}

\section{INTRODUCTION} 
As is well-known, nonzero electric dipole moment of
 elementary particles would indicate CP-violation. In the context of
 SM, there is an established source of CP-violation which is the
 famous phase of the CKM matrix. However, the contribution of this
 phase to $d_e$ is smaller than $10^{-38}~e$~cm \cite{deneutrino}
 which is too small to be probed in any foreseeable future
 \cite{forthcoming}. The phases in the neutrino mass matrix can
 also contribute to $d_e$ but their contribution is suppressed by
 fourth power of
 neutrino mass and is quite negligible: $O( 10^{-73})~e$~cm
 \cite{massNeutrinoEDM}.  Thus, detection of a nonzero $d_e$ at
 future experiments \cite{forthcoming} would open a window on new
 physics.

Another class of phenomena  that can teach us about new physics
are Lepton Flavor Violating (LFV) rare decays of charged leptons:
{\it i.e.,} $\mu \to e \gamma$, $\tau \to e \gamma$ and $\tau \to
\mu \gamma$.  It is by now established that the violation of
lepton flavor  takes place in the neutrino oscillation phenomenon;
however, if the  source of LFV is merely the neutrino mass matrix,
the rate of LFV will be extremely low \cite{LFV-Petcov} and below
the sensitivity of any search in the foreseeable future. Thus, if
the future searches record a positive signal, it will be an
indication for new physics.

 The scale of the new physics might lie at high energies (100 GeV
 or higher) but we can learn about the properties of the new
 physics by studying the indirect effects on low energy phenomena
 such as Electric Dipole Moment (EDM) and/or LFV rare decay of charged
 leptons. If there is a way to check what we have learned from the low
 energy phenomena by direct measurements at high energy labs, the
 results will be more exciting. The former can be considered as a
 guideline for the latter.

Minimal Supersymmetric Standard Model (MSSM), which is arguably
the most popular extension of the SM, introduces  several sources
for CP-violation as well as sources for LFV which can lead to
effects exceeding the present experimental bounds. The
experimental bounds on Br$(\ell_j \to \ell_i \gamma)$  and the EDM
of the elementary particles  constrain the sources of LFV and
CP-violation, respectively.  In the context of MSSM with vanishing
LFV sources, the bounds from the EDMs on the CP-violating phases
have been extensively studied in the literature (for an incomplete
list see \cite{cancelation,without cancelation,Bartl}). Although
$A_\tau$ (trilinear coupling of the staus in the soft potential)
is a LF conserving coupling, in the presence of LFV, it can affect
the properties of leptons of other generations. In particular, in
the presence of LFV, the phase of $A_\tau$ can contribute to $d_e$
at one loop level \cite{main}. In \cite{main}, the bounds  on the
LFV elements of the trilinear $A$-couplings from the stability of
vacuum was overlooked. In this paper, we take into account these
bounds and demonstrate that at certain parts of the parameter
space, these bounds reduce ambiguities in the interpretation of
results and helps us to derive conclusive bounds.

\section{SOURCES OF CP-VIOLATION AND LFV IN THE MSSM}
The phenomenology of MSSM is determined by its superpotential  and
the soft supersymmetry breaking potential. The part of the
superpotential relevant for this study is
 \be \label{superpotential}
   W_{{\rm MSSM}} =
           - Y_{i}  \widehat{{e}^c_{Ri}} \ \widehat{L}_i  \cdot \widehat{H_{d}}
-\mu\ \widehat{H_{u}}\cdot \widehat{H_{d}} \ee
 where $\widehat{L}_i$, $\widehat{H_{u}}$ and $ \widehat{H_{d}}$
 are doublets of chiral superfields respectively associated  with
 doublet $(\nu_i~e_{Li})$ and the two Higgs doublets of the
 MSSM. In the above formula,
  $\widehat{{e}^c_{Ri}}$ is
the chiral superfield associated with the  right-handed charged
lepton field $e_{Ri}$. The index $``i"$ determines the flavor. We
have written the superpotential in  the mass basis of charged
leptons ({\it i.e.,} Yukawa coupling of the charged leptons is
taken to be diagonal).   At the electroweak scale, the part of the
soft supersymmetry breaking potential  relevant for this study can
be written as \begin{eqnarray} \label{MSSMsoft}\L_{\rm soft}^{\rm
MSSM} &=&-\ 1/2 \left(  M_1 \widetilde{B}\widetilde{B}+ M_2
\widetilde{W} \widetilde{W} +{\rm H.c.} \right) \cr
&-&\left((A_{i}Y_{ i}\delta_{ij}+A_{ij}) \widetilde{e_{Ri}^c} \
\widetilde{L_{j}} \cdot H_{d} + {\rm H.c.} \right) -
\widetilde{L_{i}} ^{\dag} \ ( m_{\tilde{e}_{L }
}^{2})_{ij}\widetilde{L_{j}} - \widetilde{e_{Ri}^c }^{\dag} \
(m_{\tilde{e}_{R }}^{2})_{ij}\widetilde{e_{Rj}^c}\cr &-& \
m_{H_{u}}^{2}\ H_{u}^{\dag}\ H_{u}-\ m_{H_{d}}^{2}\ H_{d}^{\dag}\
H_{d}-(\ B_H \ H_{u}\cdot H_{d}+ {\rm H.c.}),\end{eqnarray} where
the ``$i$" and ``$j$" indices determine the flavor and
$\tilde{L}_i$ consists of $(\tilde{\nu}_i \ \tilde{e}_{Li})$.
Notice that we have divided the trilinear coupling to a flavor
diagonal  part ($A_{i}Y_{i}\delta_{ij}$) and a LFV part ($A_{ij}$
with $A_{ii}=0$). Terms involving the squarks as well as the
gluino mass term have to be added to Eq.~(\ref{MSSMsoft}). The
Hermiticity of the Lagrangian implies that $m_{H_{u}}^2$,
$m_{H_{d}}^2$ and the diagonal elements of $m_{\tilde{e}_L}^2$ and
$m_{\tilde{e}_{R }}^2$ are all real. Moreover, without loss of
generality we can rephase the fields to make  $M_2$, $B_H$ and
$Y_{ i}$ real. In such a basis, the rest of the above parameters
can in general be complex and can be considered as sources of
CP-violation  giving contributions to the EDMs.

\begin{figure}[h]\begin{center}
  \includegraphics[height= 5 cm,bb=45 125 380 350,clip=true]{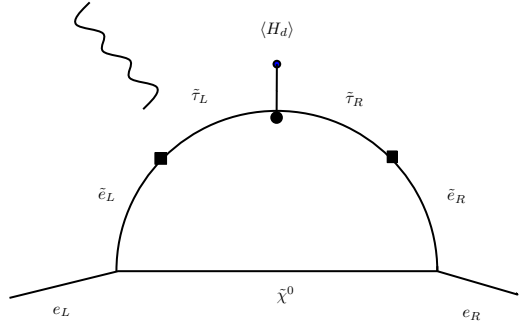}
  \caption{A neutralino exchange diagram contributing to $d_e$. The photon can attach to any of
  the $\tilde{e}_L$, $\tilde{\tau}_L$, $\tilde{\tau}_R$ or $\tilde{e}_R$ propagators. The boxes
  on the left and right sides respectively depict insertion of $(m_L^2)_{e\tau}$ and
  $(m_R^2)_{\tau e}$. The circles indicate insertion of the $A_\tau$ vertex and the vacuum expectation
  value of $H_d$.  } \label{massinsertion}
\end{center}\end{figure}

After electroweak symmetry breaking, the slepton mass terms can be
written as \begin{equation}L_{{\rm slepton}}=-\sum_{i,j}
\left(%
\begin{array}{cc}
  \widetilde{e}_{Li}^\dagger& \widetilde{e}_{Ri}^\dagger \\
\end{array}%
\right)\left(%
\begin{array}{cc}
  (m_L^2)_{ij} & (m_{LR}^{2\dag})_{ij} \\
  (m_{LR}^2)_{ij} & (m_R^2)_{ij} \\
\end{array}\right)\left(%
\begin{array}{c}
  \widetilde{e}_{Lj} \\
  \widetilde{e}_{Rj} \\
\end{array}%
\right) ,
 \end{equation}
where
\begin{equation} \label{mL}
(m_L^2)_{ij}=(m_{\widetilde{e}_L}^2)_{ij}+(m_{e}^2)_i\delta_{ij}+m_{Z}^2\cos2\beta(-\frac{1}{2}+
\sin^2\theta_W)\delta_{ij}
\end{equation}

\begin{equation} \label{mR}
(m_R^2)_{ij}=(m_{\widetilde{e}_R}^2)_{ij}+(m_{e}^2)_i\delta_{ij}-m_{Z}^2\cos2\beta\sin^2\theta_W
\delta_{ij}
\end{equation}
and
\begin{equation} \label{mLR}
(m_{LR}^2)_{ij}=m_{i}(A_{i}-\mu^*\tan\beta)\delta_{ij}+A_{ij}\langle
H_d \rangle
\end{equation}
in which $\tan \beta=\langle H_u \rangle/\langle H_d \rangle$. The
sources of LFV are $(m^2_L)_{ij}$, $(m^2_R)_{ij}$ and
$(m^2_{LR})_{ij}=A_{ij}\langle H_d \rangle$ with $i\ne j$. In the
absence of LFV, at one loop level, each of $A_\alpha$ can
contribute to the electric dipole moment of only the corresponding
charged lepton $e_\alpha$. For example,  $d_e$ receives a
significant contribution from  the phase of $A_e$ at one loop
level but if $(m^2_L)_{e\tau}=(m^2_R)_{e\tau}=A_{e\tau}=A_{\tau
e}=0$, the phase of $A_\tau$ cannot induce any contribution to
$d_e$ at one loop level. At the two loop level, imaginary
 $A_\tau$ can induce a contribution to $d_e$ but the effect is of
 course loop suppressed
\cite{two-loop,main}. When we turn on the LFV terms, imaginary
$A_\tau$ can induce a contribution to $d_e$ at one loop level
which can exceed the present bound on $d_e$ by several orders of
magnitude. The effect is demonstrated in Fig.~\ref{massinsertion}.
For illustrative purposes, in this figure, the off-diagonal
elements of $m_L^2$ and $m^2_R$ are inserted on the relevant lines
as a small perturbation.  However, to make the analysis, we use
the exact formula for $d_e$ and do not use mass insertion
approximation. The formulation can be found in the appendix of
\cite{main}.

The strong bound on Br$(\mu \to e \gamma)$ \cite{pdg} implies
strong bounds on the $e\mu$ elements of $m^2_L$, $m^2_R$ and
$A_{ij}$. Throughout this paper we set the $e \mu$ elements of
these matrices equal to zero.  There are also strong bounds on
Br($\tau \to e \gamma$) and Br($\tau \to \mu \gamma$)
\cite{Benerjee} but these bounds are about three orders of
magnitude less stringent than the bound on Br($\mu \to e \gamma$).
Each of the LFV $e\tau$ and $\mu \tau$ elements  can be sizeable
(of order of the diagonal elements) without violating the present
bounds. However if the $\tau e$ and $\tau \mu$ elements are
simultaneously present, both $\mu $ and $e$ flavor will be
violated  and  Br($\mu \to e \gamma$) can receive a contribution
exceeding the present bound on it. To avoid such a situation, we
set all the $\mu \tau$ elements equal to zero so the only sources
of LFV in the present analysis are the $e\tau$ elements.
 \section{NEW CONTRIBUTIONS TO $d_e$  IN THE PRESENCE OF LFV}
In this section, we explore the effects of $A_\tau$ on $d_e$ by
presenting figures. To draw the figures, the mass spectrum
corresponding to the $\alpha$ benchmark proposed in
\cite{NUHMbenchmark} has been chosen. However,  the mass spectrum
of the staus has been allowed  to slightly deviate from that at
the $\alpha$ benchmark. Notice that at this benchmark, the
lightest stau is considerably heavier than the lightest neutralino
so stau-neutralino coannihilation cannot play any significant role
in fixing the dark matter relic density. As a result, a slight
change of stau parameters will not dramatically affect the
cosmological predictions. We have checked for robustness of the
results and have found that the $\alpha$ benchmark is a typical
point in the parameter space that demonstrate the overall behavior
for most of the parameter space. More figures can be found in
\cite{main}.

\begin{figure}
\includegraphics[bb= 50 25 553 495, clip=true, height=4.5
in]{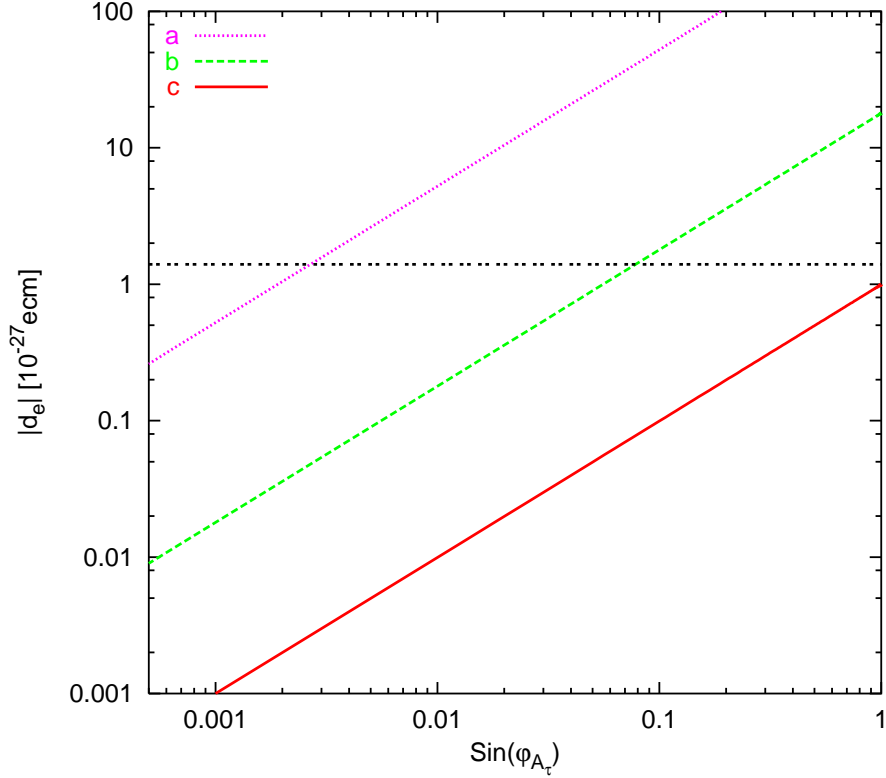} \caption{$d_e$ versus $\sin \phi_{A_\tau}$. The
input parameters correspond to the $\alpha$ benchmark proposed in
\cite{NUHMbenchmark}: $|\mu|=375$~GeV, $m_0=210$~GeV, ${\rm
M}_{1/2}=285$~GeV and $\tan \beta=10$. We have set
$|A_{\tau}|$=500~GeV. All the LFV elements of the slepton mass
matrix are set to zero except $(m_L^2)_{e \tau}$ and $(m_R^2)_{e
\tau}$. The dotted (pink) line labeled (a) corresponds to
$(m_L^2)_{e \tau}$=$3500~{\rm GeV}^2$ and $(m_R^2)_{e
\tau}$=$15000~{\rm GeV}^2$. The dashed (green) line labeled (b)
corresponds to $(m_L^2)_{e \tau}$=$50~{\rm GeV}^2$ and $(m_R^2)_{e
\tau}$=$37000~{\rm GeV}^2$. The solid (red) line labeled (c)
corresponds to $(m_L^2)_{e \tau}$=$3500~{\rm GeV}^2$ and
$(m_R^2)_{e \tau}$=$30~{\rm GeV}^2$. The horizontal doted line at
$1.4 \times 10^{-27}~e~{\rm cm}$ depicts the present experimental
 limit \cite{pdg} on $d_e$.} \label{forgotten}
\end{figure}

Fig.~\ref{forgotten} shows $d_e$ versus the sine of
$\phi_{A_\tau}$ for $A_{ij}=0$ and various values of $(m_L^2)_{e
\tau}$ and $(m_R^2)_{e \tau}$. To draw the dotted line marked with
(a), $(m_L^2)_{e \tau}$ and $(m_R^2)_{e\tau}$ are both taken to be
large. As seen from the figure, in this case the present bound on
$d_e$ (depicted by the horizontal line) puts a strong bound on the
phase of $A_\tau$. However, in the case that either
$(m_L^2)_{e\tau}$ or $(m^2_R)_{e \tau}$ is very small (as in the
case of dashed line (b) and solid line (c)), the bound is
considerably relaxed. Figure~\ref{massinsertion} demonstrates the
reason: In order for ${\rm Im}[A_\tau]$ to contribute to $d_e$,
both $(m^2_L)_{e\tau}$ and $(m_R^2)_{e\tau}$ have to be sizeable.
Suppose in the future, rare decay $\tau \to e \gamma$ is detected
which means ``some" of the $e \tau$ elements are nonzero. By
measuring only Br($\tau \to e \gamma$), one cannot determine the
ratio $(m_L^2)_{e \tau}/(m_R^2)_{e \tau}$. However, if the number
of the detected events is statistically significant, it will be
possible to derive more information by studying the angular
distribution of the final particles in the $\tau \to e \gamma$
decay \cite{kitano}.

\begin{figure}
\begin{center}
\centerline{\includegraphics[width=7.5cm]{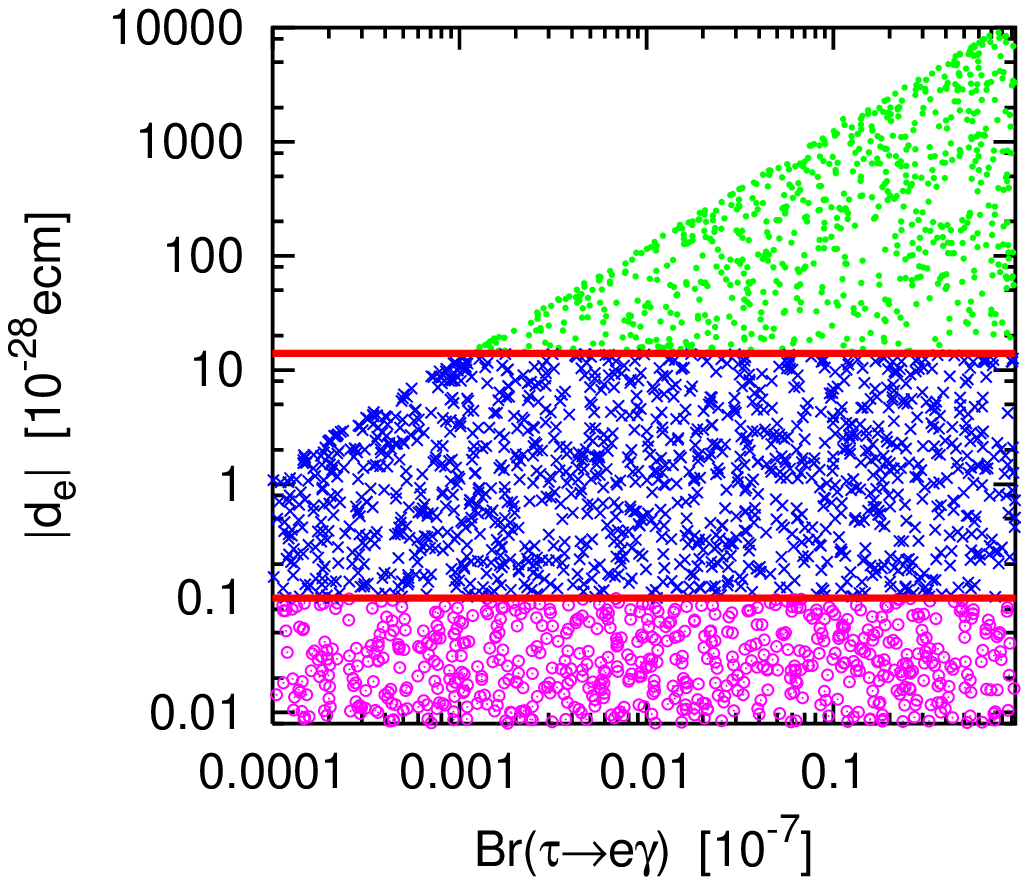}\hspace{5mm}
\includegraphics[width=7.5cm]{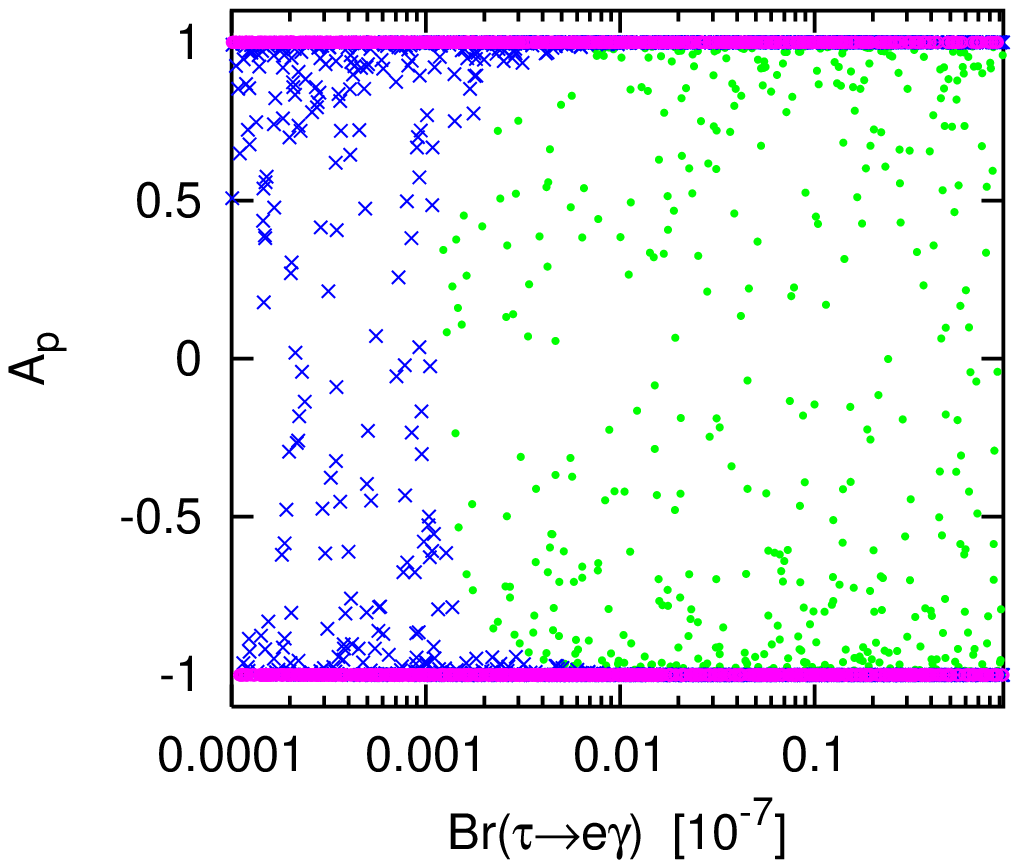}}
\centerline{\vspace{1.cm}\hspace{0.5cm}(a)\hspace{7cm}(b)}
\end{center}
\caption{ a) Scatter plot of $d_e$ versus ${\rm
Br}(\tau\rightarrow e\gamma)$. The input parameters correspond  to
the $\alpha$ benchmark proposed in \cite{NUHMbenchmark}:
$|\mu|=375$~GeV, $m_0=210$~GeV, ${\rm M}_{1/2}=285$~GeV and $\tan
\beta=10$. We have however set $\phi_{A_{\tau}}=\pi/2$ and
$|A_{\tau}|=500$~GeV. All the LFV elements of the slepton mass
matrix are set to zero except $(m^2_L)_{e\tau}$ and
$(m^2_R)_{e\tau}$ which pick up random values at a
 logarithmic scale respectively
from $(5.9\times 10^{-4}~{\rm GeV}^2,5.9 \times 10^{3}~{\rm
GeV}^2)$ and ($3.7 \times 10^{-3}~{\rm GeV}^2, 3.7\times
10^{4}~{\rm GeV}^2$).
 The horizontal line at $1.4 \times 10^{-27}~e~{\rm cm}$ depicts the present experimental
 limit \cite{pdg} and the one at $10^{-29}~e~{\rm cm}$ shows the limit that can be probed in the near future
 \cite{forthcoming}.
b) Scatter plot of $A_P$ versus ${\rm Br}(\tau\rightarrow
e\gamma)$. For each scatter point in Fig.~\ref{alphaMM}-a there is
a counterpart in Fig.~\ref{alphaMM}-b corresponding to the same
input values for the $e\tau$ elements which is shown with the same
color and symbol. Notice that points shown in pink (corresponding
to $d_e<10^{-29}~e~{\rm cm}$) all lie on the horizontal lines at
$A_p$=$\pm1$.} \label{alphaMM}
\end{figure}

\begin{figure}
\begin{center}
\centerline{\includegraphics[width=7.5cm]{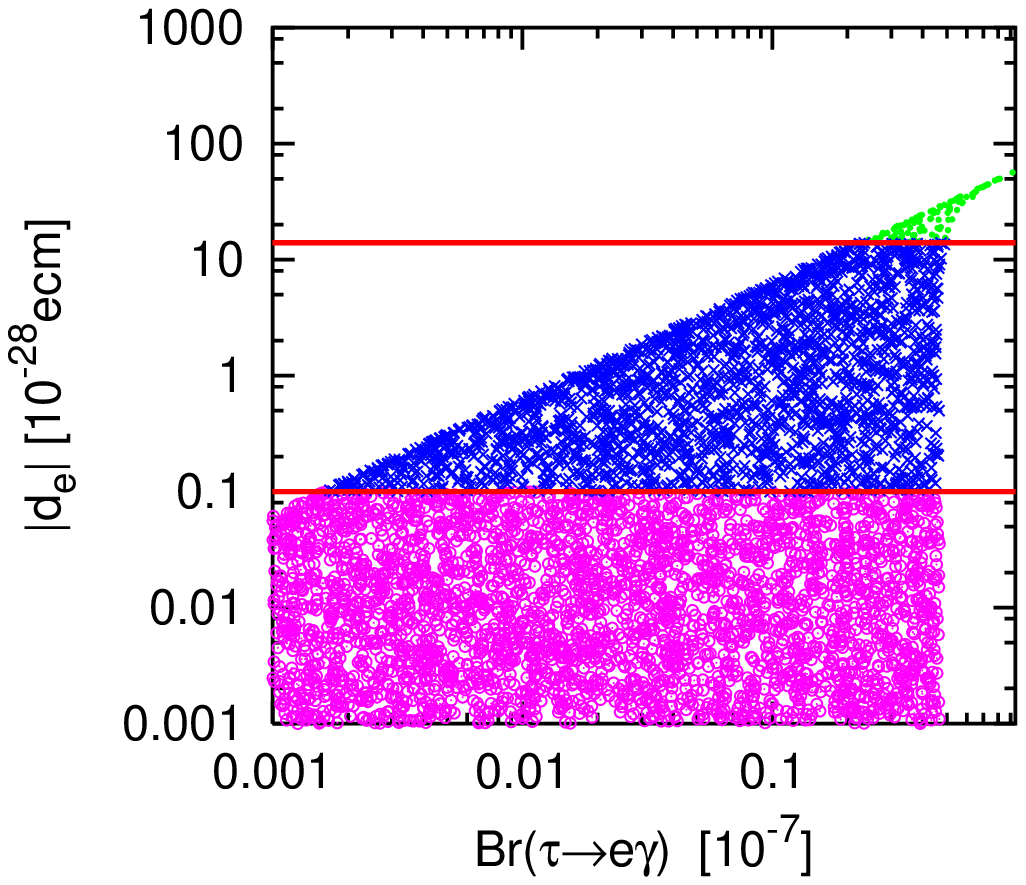}\hspace{5mm}
\includegraphics[width=7.5cm]{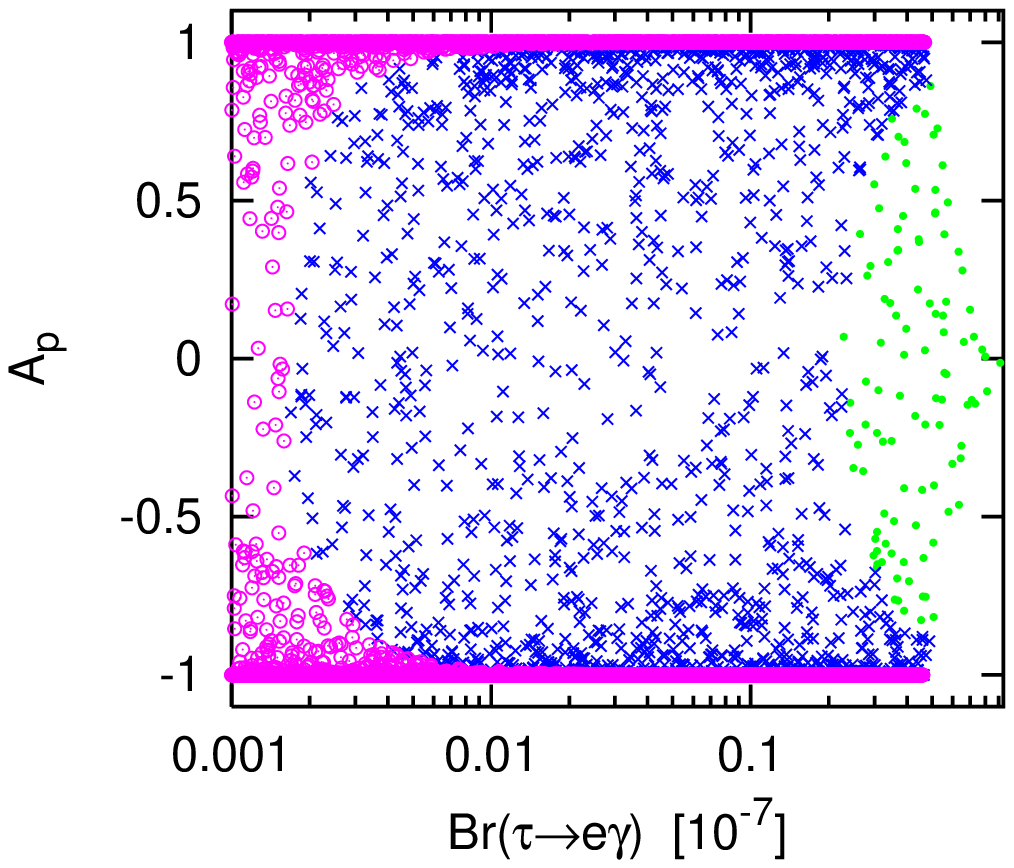}}
\centerline{\vspace{1.cm}\hspace{0.5cm}(a)\hspace{7cm}(b)}
\end{center}
\caption{Similar to Fig.~\ref{alphaMM} except that
$(m^2_L)_{e\tau}=(m^2_R)_{e\tau}=0$ and instead
$(m^2_{LR})_{e\tau}(=A_{e\tau}\langle H_d\rangle)$ and
$(m^2_{LR})_{\tau e}(=A_{ \tau e}\langle H_d\rangle)$ pick up
random values at a logarithmic scale from $(7.8 \times
10^{-4}~{\rm GeV}^2,7.8\times 10^{2}~{\rm GeV}^2)$. For each
scatter point in Fig.~\ref{alphaAA}-a there is a counterpart in
Fig.~\ref{alphaAA}-b corresponding to the same input values for
the $e\tau$ elements which is shown with the same color and
symbol.} \label{alphaAA}
\end{figure}

\begin{figure}
\begin{center}
\centerline{\includegraphics[width=7.5cm]{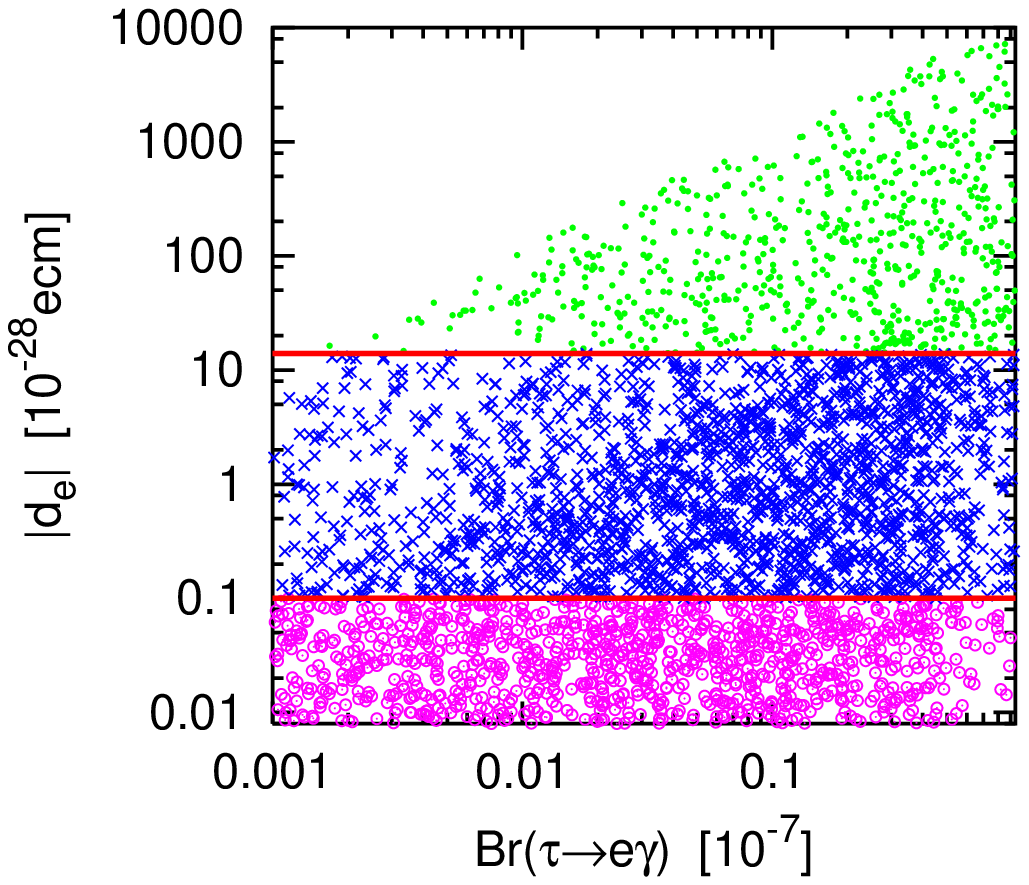}\hspace{5mm}
\includegraphics[width=7.5cm]{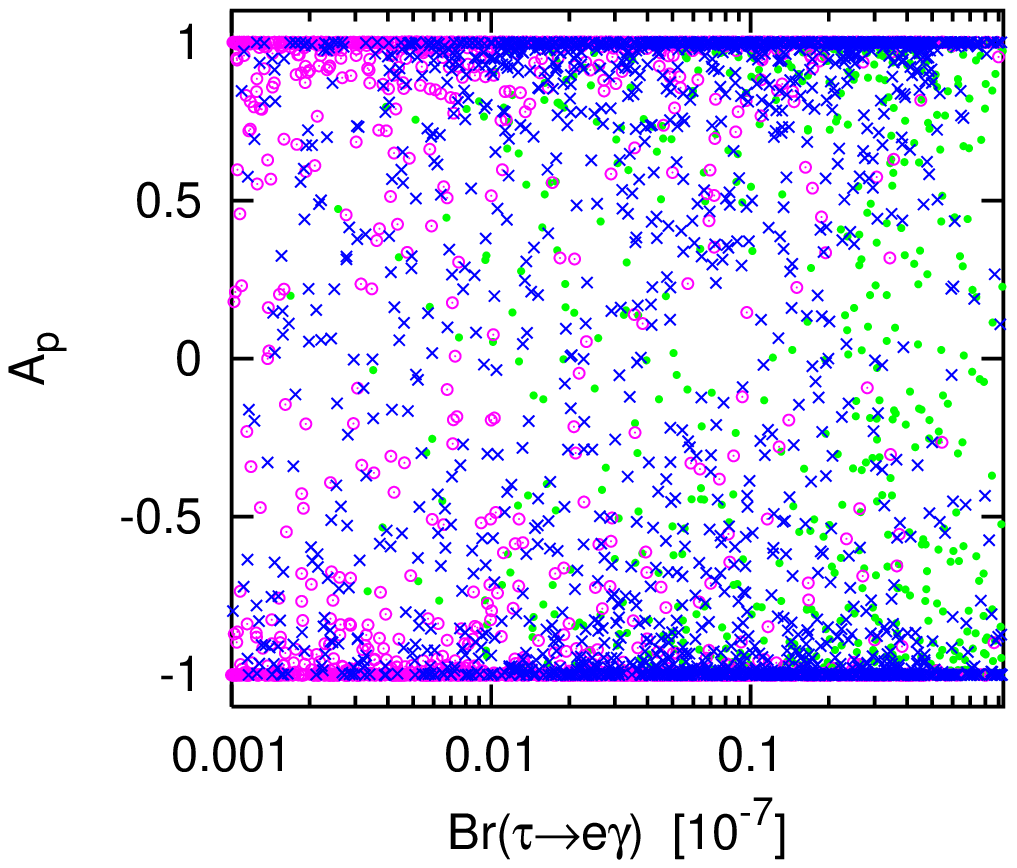}}
\centerline{\vspace{1.cm}\hspace{0.5cm}(a)\hspace{7cm}(b)}
\end{center}
\caption{Similar to Fig.~\ref{alphaMM} except that here in
addition to  $(m_L^2)_{e \tau}$ and $(m_R^2)_{e \tau}$,
$(m^2_{LR})_{e\tau}(=A_{e\tau}\langle H_d\rangle)$ and
$(m^2_{LR})_{\tau e}(=A_{ \tau e}\langle H_d\rangle)$ are also
allowed to be nonzero. The values of $(m_L^2)_{e \tau}$ and
$(m_R^2)_{e \tau}$ are randomly chosen respectively from
($5.9\times 10^{-3}~{\rm GeV}^{2},5.9 \times 10^3~{\rm GeV}^2)$
and $(3.7\times 10^{-2}~{\rm GeV}^2,3.7\times 10^4~{\rm GeV}^2 )$
at a logarithmic scale. $(m^2_{LR})_{e\tau }$ and
$(m^2_{LR})_{\tau e}$ pick up random values at a logarithmic scale
from the interval $(0.78~{\rm GeV}^2,780~{\rm GeV}^2)$. For each
scatter point in Fig.~\ref{alphaMMAA}-a there is a counterpart in
Fig.~\ref{alphaMMAA}-b corresponding to the same input values for
the $e\tau$ elements which is shown with the same color and
symbol.} \label{alphaMMAA}
\end{figure}

\begin{figure}
\begin{center}
\centerline{\includegraphics[width=7.5cm]{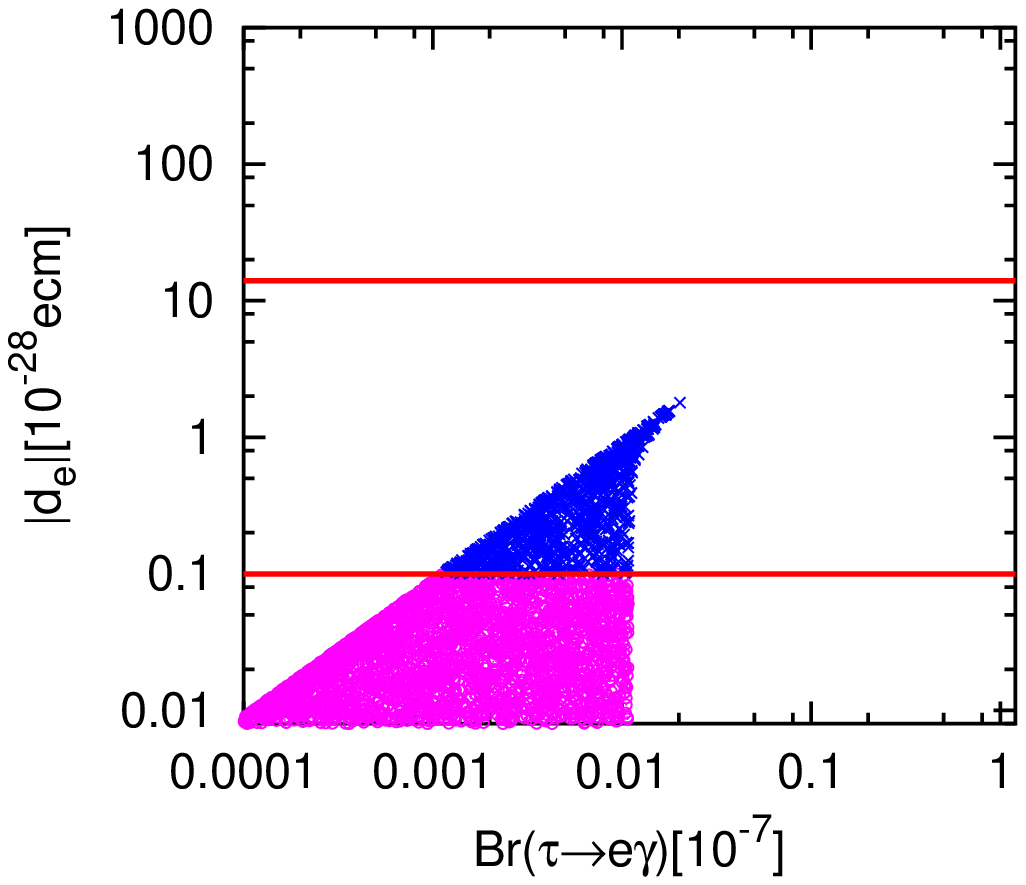}\hspace{5mm}
\includegraphics[width=7.5cm]{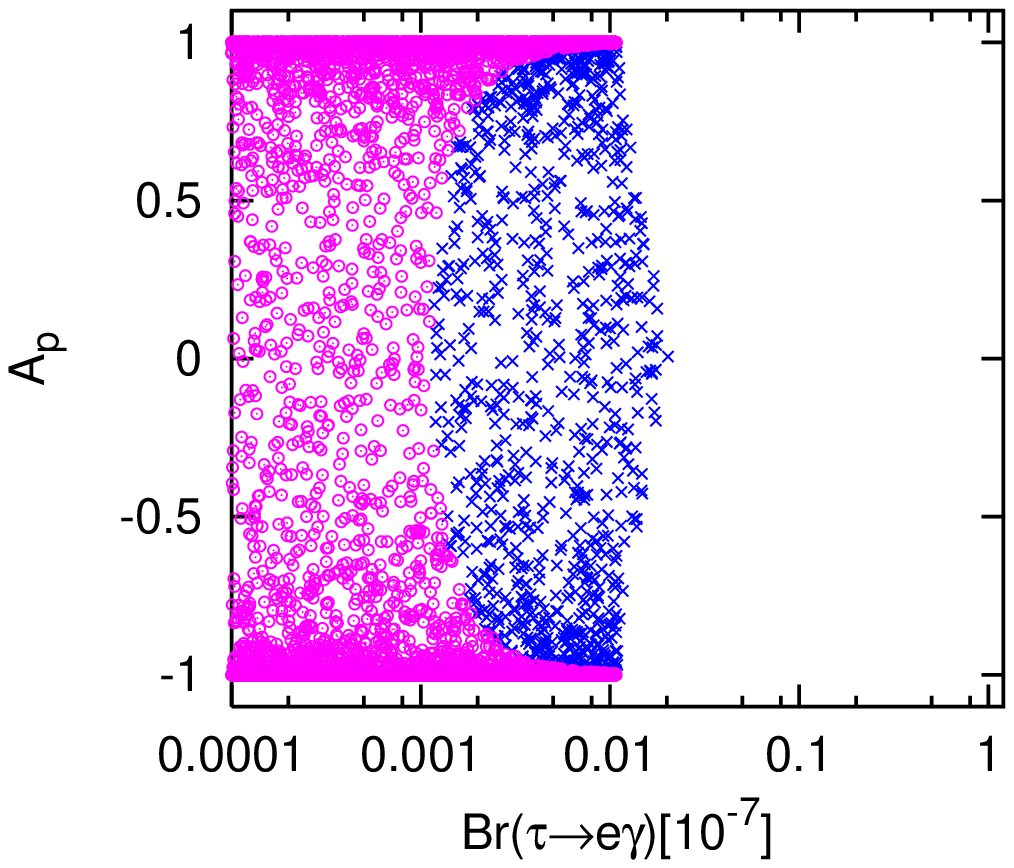}}
\centerline{\vspace{1.cm}\hspace{0.5cm}(a)\hspace{7cm}(b)}
\end{center}
\caption{a) Scatter plot of $d_e$ versus ${\rm Br}(\tau\rightarrow
e\gamma)$. The input parameters correspond  to the $\delta$
benchmark proposed in \cite{NUHMbenchmark}: $|\mu|=930$~GeV,
$m_0=500$~GeV, ${\rm M}_{1/2}=750$~GeV and $\tan \beta=10$. We
have however set $\phi_{A_{\tau}}=\pi/2$ and
$|A_{\tau}|=1800$~GeV. $(m^2_L)_{e\tau}$ and $(m^2_R)_{e\tau}$ are
zero but $(m^2_{LR})_{e\tau}$ and $(m_{LR}^2)_{\tau e}$ pick up
random values from $(0.18~{\rm GeV}^2,1.8\times 10^3~{\rm
GeV}^2)$.
 The horizontal line at $1.4 \times 10^{-27}~e~{\rm cm}$ depicts the present experimental
 limit \cite{pdg} and the one at $10^{-29}~e~{\rm cm}$ shows the limit that can be probed in the near future
 \cite{forthcoming}.
b) Scatter plot of $A_P$ versus ${\rm Br}(\tau\rightarrow
e\gamma)$. For each scatter point in Fig.~a there is a counterpart
in Fig.~b corresponding to the same input values for the $e\tau$
elements which is shown with the same color and symbol. Notice
that points shown in pink (corresponding to $d_e<10^{-29}~e~{\rm
cm}$) all lie on the horizontal lines at $A_p$=$\pm1$.}
\label{deltaAA}
\end{figure}

\begin{figure}
\begin{center}
\centerline{\includegraphics[width=7.5cm]{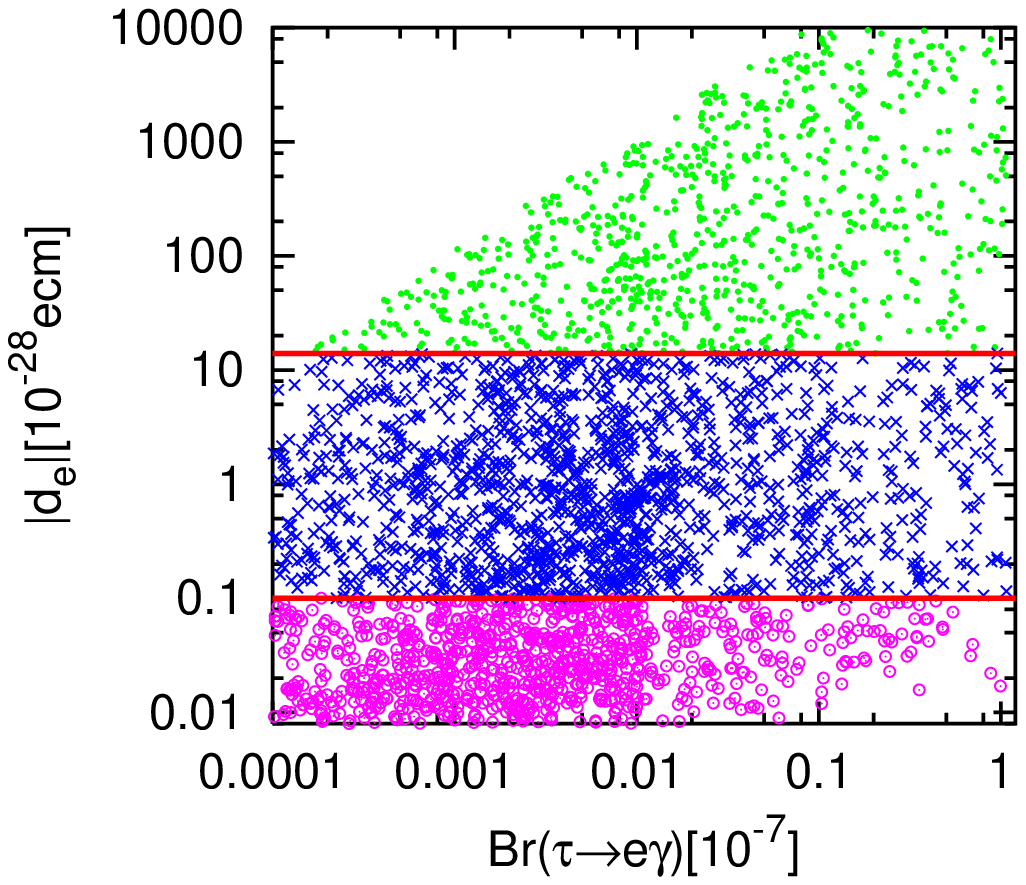}\hspace{5mm}
\includegraphics[width=7.5cm]{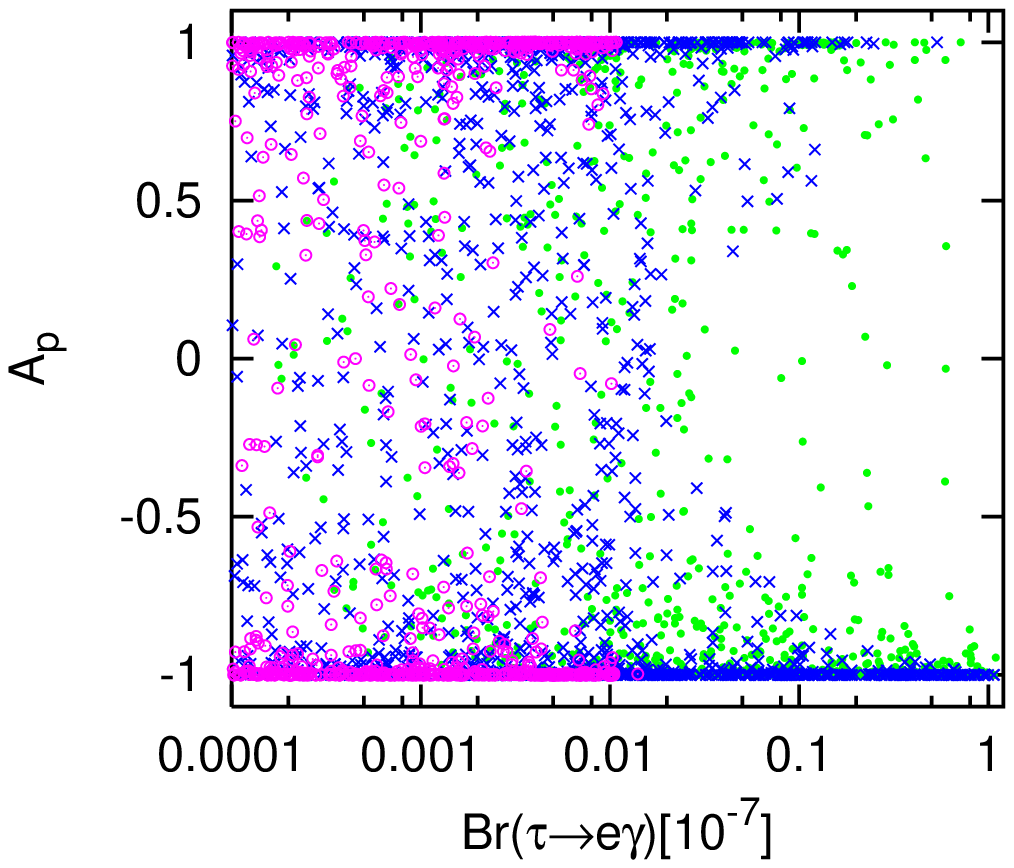}}
\centerline{\vspace{1.cm}\hspace{0.5cm}(a)\hspace{7cm}(b)}
\end{center}
\caption{a) Scatter plot of $d_e$ versus ${\rm Br}(\tau\rightarrow
e\gamma)$. The input parameters correspond  to the $\delta$
benchmark proposed in \cite{NUHMbenchmark}: $|\mu|=930$~GeV,
$m_0=500$~GeV, ${\rm M}_{1/2}=750$~GeV and $\tan \beta=10$. We
have however set $\phi_{A_{\tau}}=\pi/2$ and
$|A_{\tau}|=1800$~GeV. $(m^2_L)_{e\tau}$ and $(m^2_R)_{e\tau}$
respectively pick up random values at a
 logarithmic scale respectively
from $(0.23~{\rm GeV}^2,2.3 \times 10^{5}~{\rm GeV}^2)$ and
($0.33~{\rm GeV}^2, 3.3\times 10^{5}~{\rm GeV}^2$).
$(m^2_{LR})_{e\tau}$ and $(m_{LR}^2)_{\tau e}$ pick up random
values from $(0.18~{\rm GeV}^2,1.8\times 10^3~{\rm GeV}^2)$.
 The horizontal line at $1.4 \times 10^{-27}~e~{\rm cm}$ depicts the present experimental
 limit \cite{pdg} and the one at $10^{-29}~e~{\rm cm}$ shows the limit that can be probed in the near future
 \cite{forthcoming}.
b) Scatter plot of $A_P$ versus ${\rm Br}(\tau\rightarrow
e\gamma)$. For each scatter point in Fig.~a there is a counterpart
in Fig.~b corresponding to the same input values for the $e\tau$
elements which is shown with the same color and symbol. Notice
that points shown in pink (corresponding to $d_e<10^{-29}~e~{\rm
cm}$) all lie on the horizontal lines at $A_p$=$\pm1$.}
\label{deltaMMAA}
\end{figure}

 Following \cite{main}, let us define
\be \label{apdef}
 A_P=4\times {\int_0^1 \frac{d \Gamma(\tau\to e \gamma)}{d\cos
 \theta}d\cos \theta-\int_{-1}^0 \frac{d \Gamma(\tau\to e \gamma)}{d\cos
 \theta}d\cos \theta \over \Gamma(\tau \to e \gamma)}. \ee
 where $\theta$ is the angle that the momentum of $e$ makes with
 the spin of $\tau$. In principle, $A_P$ can be measured
 by studying the angular distribution of the final particles at an $e^-e^+$ collider such as
 a B-factory \cite{kitano}. $A_P$ is a measure of the hierarchy between left
 and right LFV elements. That is if $(m_R^2)_{e \tau} \ll
 (m_L^2)_{e\tau}$ and $A_{e\tau}\ll A_{\tau e}$, $A_P$
 converges to $-1$. In the opposite case that $(m_R^2)_{e \tau}
 \gg
 (m_L^2)_{e\tau}$ and $A_{e\tau}\gg A_{\tau e}$, $A_P$
 converges to $1$.  Figs.~(\ref{alphaMM}-\ref{deltaMMAA}) examine  the correlation
 between $A_P$ and $d_e$. To draw these plots we have assigned random values to the
$e\tau$ elements of the slepton mass matrix in the range
satisfying the present bound on Br($\tau \to e \gamma$)
\cite{Benerjee}.
  In Figs (\ref{alphaMM}-\ref{alphaMMAA}),
 we have set  $|A_\tau|=500$~GeV and assumed maximal value for the CP-violating
 phase: $\phi_{A_\tau}=\pi/2$. For the LF conserving parameters, we have
 taken   the spectrum of the $\alpha$ benchmark
 \cite{NUHMbenchmark}. Each pair of the scatter plots
 shown in Figs.~(\ref{alphaMM}-\ref{alphaMMAA})  corresponds to
 different configurations of the
 $e\tau$ elements. To draw each pair, we have assigned random values (from a logarithmic
 scale) to  various $e\tau$ elements. We have then  calculated the corresponding
 values of Br($\tau \to e \gamma$), $|d_e|$ and $A_P$ and have depicted
 the corresponding scatter points with the same color and symbol in figures
 (a) and (b). The horizontal lines  at $1.4
 \times 10^{-27}~e$~cm and $10^{-29}~e$~cm respectively show the
 present bound \cite{pdg} and the reach of the forthcoming
 experiments \cite{forthcoming}.
 As seen from the figures, for a given value of Br($\tau \to e
 \gamma$), $d_e$ cannot exceed a certain value.

In the case of Fig.~\ref{alphaMM},  $A_{e\tau}$ and $A_{\tau e}$
are set equal to zero. As seen from the figure, for a significant
portion of the parameter space, $d_e$ lies above the present bound
(the points shown with green (light grey) dots).  The scatter
points depicted by  pink circle, which appear in
Fig.~\ref{alphaMM}-b as two pink horizontal lines at $A_P=\pm 1$,
correspond to $d_e<10^{-29}~e~{\rm cm}$. From Fig.~\ref{alphaMM}-b
we conclude that for $A_{e \tau}=A_{\tau e}=0$, the bound on $d_e$
can be satisfied if either ${\rm Br}(\tau \to e \gamma)$ is very
small (which means that all the LFV masses are very small) or
$A_P$ is close to $\pm 1$ (meaning that there is a hierarchy
between the LFV elements). In other words within this scenario, if
future searches find $5 \times 10^{-10}<{\rm Br}(\tau \to e
\gamma)$ and $-0.9<A_P<0.9$, the bound on $d_e$ should be
interpreted either as a bound on $\phi_{A_\tau}$ or as an
indication for a cancelation between different contributions from
$\phi_{A_\tau}$ and other possible CP-violating phases.

To draw Fig.~\ref{alphaAA}, $(m_L^2)_{e \tau}$ and $(m_R^2)_{e
\tau}$ are set equal to zero and instead random values within a
range are assigned to $(m_{LR}^2)_{e \tau}$ and $(m_{LR}^2)_{\tau
e}$. The upper limit of the range  ({\it i.e.,} 780~GeV$^2$)
saturate the constraints from the Unbounded From Below (UFB)
consideration \cite{Savas}. Notice that these bounds on
$(m_{LR}^2)_{e \tau}$ and $(m_{LR}^2)_{\tau e}$ imply a
``theoretical" bound on Br($\tau \to e \gamma$).  The scatter
points at the tilted peak  with highest $d_e$ and Br($\tau \to e
\gamma$) correspond to the cases that both $(m_{LR}^2)_{e\tau}$
and $(m_{LR}^2)_{\tau e}$ are close to the upper limit. Notice
that a correlation between $d_e$, $A_P$ and ${\rm Br}(\tau \to e
\gamma)$ similar to that in the case of Fig.~\ref{alphaMM}
emerges. That is the points marked with green dots (corresponding
to $d_e>$$1.4\times10^{-27}~e~{\rm cm})$, with blue ``$\times$"
(corresponding to $10^{-29}<d_e<1.4\times 10^{-27}~e~{\rm cm}$)
and with pink circles (corresponding to $d_e<10^{-29}~e~{\rm cm}$)
are respectively scattered  from right to left. Notice however
that in contrast to Fig.~\ref{alphaMM}-b, Fig.~\ref{alphaAA}-b
includes scatter points with $-0.9<A_P<0.9$ and ${\rm Br}(\tau \to
e \gamma)\sim 10^{-8}$ that satisfy the present bound on $d_e$
(the points marked with ``$\times$" in the plot). In
Fig.~\ref{deltaAA}, we have repeated the same analysis with the
$\delta$ benchmark \cite{NUHMbenchmark}. In the case of the
$\delta$ benchmark, the constraint from the UFB is so stringent
that for all scatter points $d_e<2 \times 10^{-28}~e$~cm and Br($
\tau \to e \gamma)<2\times 10^{-9}$.


In Fig.~\ref{alphaMMAA},  $(m_L^2)_{e\tau}$, $(m_R^2)_{e \tau}$,
$(m_{LR}^2)_{e\tau}$ and $(m_{LR}^2)_{\tau e}$ all take nonzero
random values. Fig.~\ref{alphaMMAA}-a contains features of both
Figs.~\ref{alphaMM}-a and \ref{alphaAA}-a. The significant point
is that setting all the $e \tau$ mass elements nonzero, the
correlation among $A_P$, $d_e$ and ${\rm Br}(\tau \to e \gamma)$
becomes weaker. That is, unlike Figs.~\ref{alphaMM}-b and
\ref{alphaAA}-b, Fig.~\ref{alphaMMAA}-b contains points below the
sensitivity limit of the forthcoming $d_e$ searches (points
depicted with pink circles with $d_e<10^{-29}~e$~cm)  for which
${\rm Br}(\tau \to e \gamma)> 10^{-8}$ and $-0.9<A_P<0.9$. This
can be explained as follows. At scatter points for which \be
[(m_{LR}^2)_{e\tau},(m_L^2)_{e \tau} \ll (m_{LR}^2)_{\tau
e},(m_R^2)_{e\tau}] \ \ {\rm or} \ \
[(m_{LR}^2)_{e\tau},(m_L^2)_{e \tau} \gg (m_{LR}^2)_{\tau
e},(m_R^2)_{e\tau}],\label{configuration}\ee   $ A_L$ can be of
order of $A_R$ which yields $-0.9<A_P<0.9$ but despite sizeable
$\phi_{A_\tau}$, $d_e$ is still small. Pink circles lying in the
region ${\rm Br}(\tau \to e \gamma)>  10^{-9}$ and $-0.9<A_P<0.9$
correspond to such configurations. As a result, without
independent knowledge of the ratios of LFV elements, we cannot
derive any conclusive bound on $\phi_{A_\tau}$.
The fraction of the scatter points with $d_e<10^{-29}~e$~cm (pink
circles) lying in the region with $-0.9<A_P<0.9$ and $10^{-8}<{\rm
Br}(\tau \to e \gamma)$ strongly depends on the choice of the
range and scale of random pick up of the LFV input. For example,
had we chosen the lower limit of the range of $(m_L^2)_{e\tau}$
and $(m_R^2)_{e\tau}$ two orders of magnitude higher [{\it i.e.,}
$(m_L^2)_{e\tau}\in (0.59~{\rm GeV}^2, 5900~{\rm GeV}^2)$ and
$(m_R^2)_{e \tau} \in (3.7~{\rm GeV}^2,37000~{\rm GeV}^2)$ instead
of $(m_L^2)_{e\tau}\in (0.0059~{\rm GeV}^2, 5900~{\rm GeV}^2)$ and
$(m_R^2)_{e \tau} \in (0.037~{\rm GeV}^2,37000~{\rm GeV}^2)$], no
pink circles would  have in practice   appeared in this region.
This is understandable because for a constant number of the
scatter points, decreasing the lower limit of $(m_L^2)_{e \tau}$
and $(m_R^2)_{e\tau}$ increases the weight of the region for which
condition in Eq.~(\ref{configuration}) is satisfied.

 If by some theoretical
consideration we exclude the possibility of conditions
(\ref{configuration}), the correlation between $A_P$ and $d_e$ is
maintained so, for Br($\tau \to e
\gamma)\stackrel{>}{\sim}10^{-8}$ and $-0.9<A_P<0.9$, the present
bound on $d_e$ can be interpreted as a strong bound on
$\phi_{A_\tau}$ \cite{main}. For example within the scenario
described in \cite{vives} which relates all the LFV elements to
the Yukawa couplings,  conditions (\ref{configuration}) cannot be
fulfilled. Moreover, in some parts of the parameter space, by
combining information from different observables with the UFB
bounds on the LFV elements of  $m^2_{LR}$, we can exclude the
possibility of Eq.~(\ref{configuration}). This is demonstrated in
Figs.~\ref{deltaAA} and \ref{deltaMMAA}. As seen from
Fig.~\ref{deltaAA}, at the $\delta$ benchmark, the bounds from UFB
exclude the possibility of a contribution from $(m_L^2)_{e\tau}$
and $(m_R^2)_{e\tau}$  to Br$(\tau \to e \gamma)$ at the level of
$O(10^{-8})$. Thus, if the $\delta$ benchmark is established at
LHC and Br($\tau \to e \gamma$) turns out to be of order of
$10^{-8}$, we will conclude that the contributions comes from
$(m_L^2)_{e\tau}$ and/or $(m_R^2)_{e\tau}$. Moreover if
observation shows that $-0.9<A_P<0.9$, we will conclude that the
contributions of $(m_L^2)_{e\tau}$ and $(m_R^2)_{e\tau}$ are
comparable so the conditions in Eq.~(\ref{configuration}) cannot
be fulfilled. In fact, Fig.~\ref{deltaMMAA} shows that at the
$\delta$ benchmark, the correlation between $A_P$ and $d_e$ is
maintained even when all the $e\tau$ elements pick up nonzero
values within the allowed region.
 Fig.~\ref{deltaMMAA} shows that for
Br($\tau \to e \gamma)>10^{-8}$ and $-0.5<A_P<0.5$,
$\phi_{A_\tau}=\pi/2$ yields $d_e$ higher than the present bound:
$d_e>1.4 \times 10^{-27}~e$~cm. Moreover for Br($\tau \to e
\gamma)>10^{-9}$ and $-0.9<A_P<0.9$, $\phi_{A_\tau}=\pi/2$ yields
$d_e$ detectable in forthcoming experiments: $d_e>10^{-29}~e$~cm.

\section{CONCLUSIONS AND OUTLOOK}
We have shown that in the presence of the $e\tau$ LFV elements,
the phase of $A_\tau$, $\phi_{A_\tau}$, can contribute to $d_e$ at
one loop level. For values of Br($\tau \to e \gamma)$ close to the
present experimental bounds, the contribution of $\phi_{A_\tau}$
to $d_e$ can exceed the experimental bound on $d_e$ by several
orders of magnitude. We have found that even if Br($\tau \to e
\gamma$) is three orders of magnitude below the present bound, the
contribution to $d_e$ can still exceed the present bound on $d_e$.
The effect of $\phi_{A_\tau}$ on $d_e$ strongly depends on the
ratios of the LFV slepton masses $(m_L^2)_{e\tau}/(m_R^2)_{e\tau}$
and $(m_{LR}^2)_{e\tau}/(m_{LR}^2)_{\tau e}$. In other words, for
a given value of ${\rm Br}(\tau \to e\gamma)$ and $\phi_{A_\tau}$,
$|d_e|$ can take any value between zero and a maximum which
depends on the value of ${\rm Br}(\tau \to e\gamma)$ and
$\phi_{A_\tau}$ [see Figs.~(\ref{alphaMM}-a)-(\ref{deltaMMAA}-a)].
We have shown that for the specific case that $(m_{LR}^2)_{e
\tau}=(m_{LR}^2)_{\tau e}=0$ (see Fig.~\ref{alphaMM}-b) or
$(m_{L}^2)_{e\tau}=(m_{R}^2)_{e\tau}=0$ [see
Fig.~\ref{alphaAA}-b], by measuring the asymmetry  $A_P$ defined
in Eq.~(\ref{apdef}) we can solve this ambiguity. However, in the
general case that all the $e\tau$ elements are nonzero, as shown
in Fig.~\ref{alphaMMAA}, the correlation between $A_P$ and $d_e$
becomes weaker and to solve the ambiguity, extra information is
needed.

 Let us suppose that
Br($\tau \to e\gamma$) turns out to be close to the present bound
({\it i.e.,} Br($\tau \to e \gamma)>10^{-8}$) and moreover let us
suppose  $A_P$ is measured and found to be $-0.9<A_P<0.9$.
Excluding the possibility of a fine tuned cancelation between the
contributions of different phases, two possibilities emerge: 1)
$\phi_{A_\tau}$ is smaller than $O(0.005)$; 2) $\phi_{A_\tau}$ is
large but one of the conditions in Eq.~(\ref{configuration}) is
fulfilled. To derive a conclusive bound on $\phi_{A_\tau}$, the
second possibility has to be excluded. We show that at some parts
of the parameter space such as the $\delta$ benchmark, the second
possibility is excluded by bounds on $(m_{LR}^2)_{e\tau}$ and
$(m_{LR}^2)_{\tau e}$ from the UFB consideration.

In summary,  combining the information on $d_e$ and LFV $\tau$
decay modes gives invaluable information on $\phi_{A_\tau}$. In
certain parts of the parameter space ({\it e.g.,} the $\delta$
benchmark), by studying these observables, we can constrain
$\phi_{A_\tau}$ however in other parts ({\it e.g.,} the $\alpha$
benchmark) drawing conclusive bounds on $\phi_{A_\tau}$ is not
possible.  In the latter case, this method cannot replace the
direct measurement of $\phi_{A_\tau}$ at ILC. On the other hand,
direct measurement of $\phi_{A_\tau}$ at ILC can help us to
resolve the degeneracies in the pattern of the LFV elements.


\begin{acknowledgments}
Y.F. would like to thank the organizers of ICHEP 08 for giving her
the opportunity to present this talk.
\end{acknowledgments}



\begin{thebibliography}{99} 
\bibitem{deneutrino}
  F.~Hoogeveen,
  Nucl.\ Phys.\  B {\bf 341} (1990) 322.
  M.~E.~Pospelov and I.~B.~Khriplovich,
  Sov.\ J.\ Nucl.\ Phys.\  {\bf 53} (1991) 638
  [Yad.\ Fiz.\  {\bf 53} (1991) 1030];
  M.~J.~Booth,
  arXiv:hep-ph/9301293.


\bibitem{forthcoming}
  D.~Kawall, F.~Bay, S.~Bickman, Y.~Jiang and D.~DeMille,
  AIP Conf.\ Proc.\  {\bf 698} (2004) 192;
  D.~Kawall, F.~Bay, S.~Bickman, Y.~Jiang and D.~DeMille,
  Phys.\ Rev.\ Lett.\  {\bf 92} (2004) 133007
  [arXiv:hep-ex/0309079];
  S.~K.~Lamoreaux,
  Phys.\ Rev.\ D {\bf 66} (2002) 010001
  arXiv:nucl-ex/0109014.

\bibitem{massNeutrinoEDM}
 A.~de Gouvea and S.~Gopalakrishna,
  Phys.\ Rev.\  D {\bf 72}, 093008 (2005)
  [arXiv:hep-ph/0508148].

\bibitem{LFV-Petcov}
   S.~T.~Petcov,
  Sov.\ J.\ Nucl.\ Phys.\  {\bf 25}, 340 (1977)
  [Yad.\ Fiz.\  {\bf 25}, 641 (1977\ ERRAT,25,698.1977\
  ERRAT,25,1336.1977)];
S.~M.~Bilenky, S.~T.~Petcov and B.~Pontecorvo,
  Phys.\ Lett.\  B {\bf 67}, 309 (1977);
G.~Altarelli, L.~Baulieu, N.~Cabibbo, L.~Maiani and R.~Petronzio,
  Nucl.\ Phys.\  B {\bf 125}, 285 (1977)
  [Erratum-ibid.\  B {\bf 130}, 516 (1977)].





\bibitem{cancelation}
 K.~A.~Olive, M.~Pospelov, A.~Ritz and Y.~Santoso,
  Phys.\ Rev.\ D {\bf 72} (2005) 075001
  [arXiv:hep-ph/0506106];
  S.~Abel, S.~Khalil and O.~Lebedev,
  Nucl.\ Phys.\ B {\bf 606} (2001) 151
  [arXiv:hep-ph/0103320];
  T.~Falk {\it et al.,}
  Nucl.\ Phys.\ B {\bf 560} (1999) 3
  [arXiv:hep-ph/9904393];
  A.~Afanasev, C.~E.~Carlson and C.~Wahlquist,
  Phys.\ Rev.\ D {\bf 61} (2000) 034014
  [arXiv:hep-ph/9903493];
  T.~Ibrahim and P.~Nath,
  Phys.\ Lett.\ B {\bf 418} (1998) 98
  [arXiv:hep-ph/9707409];
  M.~Brhlik, G.~J.~Good and G.~L.~Kane,
  Phys.\ Rev.\ D {\bf 59} (1999) 115004
  [arXiv:hep-ph/9810457];
  A.~Bartl {\it et al.},
  Phys.\ Rev.\ D {\bf 60} (1999) 073003
  [arXiv:hep-ph/9903402];
  T.~Falk, K.~A.~Olive, M.~Pospelov and R.~Roiban,
  Nucl.\ Phys.\ B {\bf 560} (1999) 3
  [arXiv:hep-ph/9904393];
 S.~Yaser Ayazi and Y.~Farzan,
  Phys.\ Rev.\  D {\bf 74}, 055008 (2006)
  [arXiv:hep-ph/0605272];
S.~Y.~Ayazi,
{\it In the Proceedings of IPM School and Conference on Lepton and
Hadron Physics (IPM-LHP06), Tehran, Iran, 15-20 May 2006, pp 0004}
  [arXiv:hep-ph/0611056].


\bibitem{without cancelation}
  P.~Nath,
  Phys.\ Rev.\ Lett.\  {\bf 66} (1991) 2565 ;
  Y.~Kizukuri and N.~Oshimo,
  Phys.\ Rev.\ D {\bf 46} (1992) 3025;
  V.~A.~Kuzmin, V.~A.~Rubakov and M.~E.~Shaposhnikov,
  Phys.\ Lett.\ B {\bf 155}, 36 (1985);
  V.~Cirigliano, S.~Profumo and M.~J.~Ramsey-Musolf,
  JHEP {\bf 0607}, 002 (2006)
  [arXiv:hep-ph/0603246]
  K.~A.~Olive, M.~Pospelov, A.~Ritz and Y.~Santoso,
  Phys.\ Rev.\ D {\bf 72} (2005) 075001
  [arXiv:hep-ph/0506106];
  T.~Falk and K.~A.~Olive,
  Phys.\ Lett.\ B {\bf 375} (1996) 196
  [arXiv:hep-ph/9602299];
  T.~Ibrahim and P.~Nath,
  arXiv:hep-ph/0210251.

\bibitem{Bartl}
  A.~Bartl, W.~Majerotto, W.~Porod and D.~Wyler,
  Phys.\ Rev.\ D {\bf 68} (2003) 053005
  [arXiv:hep-ph/0306050];
  W.~Porod,
{\it Prepared for International Workshop on Astroparticle and
High-Energy Physics (AHEP-2003), Valencia, Spain, 14-18 Oct 2003}.

\bibitem{main}
 S.~Y.~Ayazi and Y.~Farzan,
  JHEP {\bf 0706}, 013 (2007)
  [arXiv:hep-ph/0702149].

\bibitem{two-loop}
  D.~Chang, W.~Y.~Keung and A.~Pilaftsis,
  Phys.\ Rev.\ Lett.\  {\bf 82} (1999) 900
  [Erratum-ibid.\  {\bf 83} (1999) 3972]
  [arXiv:hep-ph/9811202].
  \bibitem{pdg}
 W.-M. Yao {\it et al.}, J. Phys. G 33 (2006) 1.
\bibitem{Benerjee}
  S.~Banerjee,
  Nucl.\ Phys.\ Proc.\ Suppl.\  {\bf 169} (2007) 199
  [arXiv:hep-ex/0702017].




\bibitem{NUHMbenchmark}
    A.~De Roeck, J.~R.~Ellis, F.~Gianotti, F.~Moortgat, K.~A.~Olive and L.~Pape,
  Eur.\ Phys.\ J.\  C {\bf 49}, 1041 (2007)
  [arXiv:hep-ph/0508198].



 \bibitem{kitano}
   R.~Kitano and Y.~Okada,
  Phys.\ Rev.\ D {\bf 63} (2001) 113003
  [arXiv:hep-ph/0012040].
\bibitem{Savas}
 J.~A.~Casas and S.~Dimopoulos,
  Phys.\ Lett.\  B {\bf 387}, 107 (1996)
  [arXiv:hep-ph/9606237].


  \bibitem{vives}
    L.~Calibbi, J.~Jones-Perez and O.~Vives,
  arXiv:0804.4620 [hep-ph].

\end{thebibliography}
\end{document}